\documentclass[twocolumn]{aastex62}
\usepackage{amsmath}

\newcommand{\di}[2]{\frac{d #1}{d #2}}

\newcommand{\citepl}[1]{\citeauthor{#1} \citeyear{#1}}

\graphicspath{{./}{figures/}}

\begin{document}
	
	\title{Atmospheric Evolution on Low-gravity Waterworlds}

	\author[0000-0002-0791-1814]{Constantin W. Arnscheidt}
	\affiliation{John A. Paulson School of Engineering and Applied Sciences, Harvard University}
	\affiliation{Department of Physics, Harvard University}
	\affiliation{Department of Earth, Atmospheric, and Planetary Sciences, Massachusetts Institute of Technology}
	
	\author[0000-0003-1127-8334]{Robin D. Wordsworth}
	\affiliation{John A. Paulson School of Engineering and Applied Sciences, Harvard University}
	\affiliation{Department of Earth and Planetary Sciences, Harvard University}
	
	\author[0000-0001-7758-4110]{Feng Ding}
	\affiliation{John A. Paulson School of Engineering and Applied Sciences, Harvard University}
	
	
	
	\begin{abstract}
		
		Low-gravity waterworlds ($M\lesssim 0.1 M_{\oplus}$) are of interest for their potential habitability. The weakly bound atmospheres of such worlds have proportionally larger radiative surfaces and are more susceptible to escape. We conduct a unified investigation into these phenomena, combining analytical energy balance and hydrodynamic escape with line-by-line radiative transfer calculations.  Because outgoing radiation is forced to increase with surface temperature by the expansion of the radiative surface, we find that these worlds do not experience a runaway greenhouse. Furthermore, we show that a long-lived liquid water habitable zone is possible for low-gravity waterworlds of sufficient mass. Its inner edge is set by the rate of atmospheric escape because a short-lived atmosphere limits the time available for life to evolve. In describing the physics of the parameter space transition from ``planet-like" to ``comet-like", our model produces a lower bound for habitability in terms of gravity. These results provide valuable insights in the context of the ongoing hunt for habitable exoplanets and exomoons.

	\end{abstract}
	
	
	
	\section{Introduction} \label{sec:intro}
	
	Earth is the classic example of a terrestrial-mass, water-rich body with both an atmosphere and surface liquid H$_2$O. Low-mass water-rich bodies (i.e. comets) have neither: the gravity is not strong enough to bind an atmosphere, and when they are heated their water ice is sublimated without any significant amounts of liquid water appearing on their surfaces. What happens at intermediate masses? 
	
	The evolution of low-gravity waterworlds differs from that of larger worlds primarily via two particular phenomena. First, the atmospheres of low-gravity bodies will expand outwards significantly as a function of surface temperature. The effective radiative surfaces of these atmospheres thus gain area relative to the surface, increasing both outgoing and absorbed radiation. Previously, it has been shown that this effect is already important for worlds of mass $\sim 0.1 M_{\oplus}$ \citep{goldblatt15}. Second, the atmospheres of low-gravity worlds are extremely vulnerable to hydrodynamic escape via supersonic pressure-driven bulk outflow \citep{pierrehumbert10, catling17, lehmer17, zahnle17}. 
	
	\cite{kuramoto94} studied these two phenomena together in the context of giant icy satellite accretion. That study used gray radiative transfer and focused on the implications for compositional evolution. Here, we present the first unified study of these phenomena and their implications for long-term atmospheric evolution and habitability over a wide range of starting parameters. To achieve this, we combine a line-by-line radiative-convective climate model (Section \ref{sec:rt}) with energy balance (Section \ref{sec:eb}) and hydrodynamic escape of the atmosphere (Section \ref{sec:hde}).
	
	The habitability of low-gravity waterworlds has previously been discussed in a range of contexts. Examples of interest in our solar system include the icy Jovian moons Europa, Ganymede, and Callisto \citep{reynolds83, chyba01, chelaflores10, grasset13}, which all have masses below $0.03 M_{\oplus}$. For terrestrial-mass worlds, habitability is characterized by a circumstellar ``habitable zone", commonly defined as the range in orbital distance within which the world could maintain surface liquid water. \citep{huang59,kasting93,kopparapu13,seager13,ramirez18}. The inner edge of the habitable zone is dependent on gravity even for near-terrestrial mass objects, albeit due to two competing mechanisms: as gravity decreases, a decreasing runaway greenhouse emission limit tends to move it further from the host star \citep{pierrehumbert10, kopparapu14}, and radiative surface expansion tends to move it closer \citep{goldblatt15}. However, because neither comets nor gas giants are habitable according to the criterion of surface liquid water \citep{kasting93,ramirez18}, there must ultimately also exist habitability boundaries in terms of gravity. The lower gravity boundary marks the transition between worlds that are effectively \textit{``planet-like"} (habitable given the correct stellar flux), and \textit{``comet-like"} (never habitable regardless of stellar flux). The model developed in this study provides a means to determine this boundary (Section \ref{sec:planetcomet}), and shows what happens as it is approached. 
	
	In Section \ref{sec:norg}, we show that atmospheric expansion suppresses the runaway greenhouse and that escape rates determine the inner edge of the habitable zone for bodies of mass $\lesssim 0.1 M_{\oplus}$.
	In Section \ref{sec:hz}, we show that a long-lived liquid water habitable zone is possible. In Section \ref{sec:hysteresis}, we show that the ice-albedo feedback works against stellar flux-driven deglaciations to long-lived states. We discuss these results in Section \ref{sec:disc} and conclude in Section \ref{sec:conc}.
	
	\section{Methods}
	\subsection{Starting assumptions}
	We assume that the low-gravity waterworld has a pure water vapor atmosphere and a water reservoir fixed at $40\%$ of the planet's total mass. For a given surface temperature, the atmospheric water content and temperature profile are set by the saturation vapor pressure. The water content of the atmosphere remains orders of magnitude below the world's total water content in even the most extreme cases considered. We assume spherical symmetry; this is required by both the radiative transfer model and the hydrodynamic escape formulation. Our model, and therefore the ``habitable zone" we consider, neglects the potential effects of CO$_2$ cycling \citep{abbot12,ramirezlevi18}, ocean chemistry \citep{wordsworth13,kite18}, and three-dimensional dynamics \citep{pierrehumbert16,ding18}.
	
	It is convenient to parameterize the mass and radius of the waterworld in terms of a single quantity, surface gravity $g$. This is accomplished using the following scaling relation, accurate for waterworlds in the range $10^{-2} M_{\oplus}<M<M_{\oplus}$ \citep{sotin07}: 
	\begin{equation}
	\frac{R}{R_{\oplus}}=1.258\left(\frac{M}{M_{\oplus}}\right)^{0.302}
	\label{sotin}
	\end{equation}
	
	\subsection{Energy balance}
	\label{sec:eb}
\begin{figure*}[!ht]
		\begin{center}
			\includegraphics[width=0.5\linewidth]{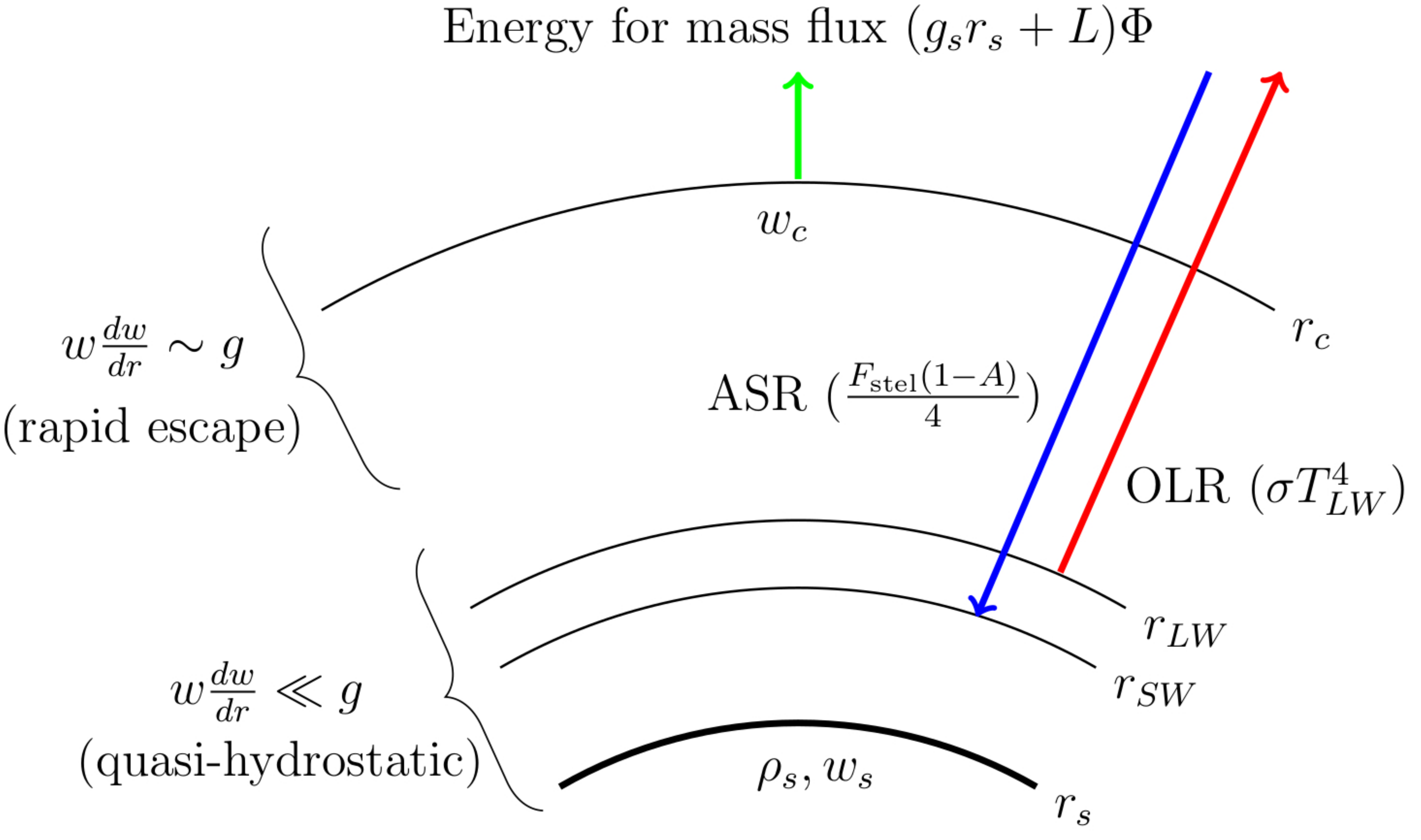}
			\caption{Schematic of our model. Absorbed shortwave radiation (ASR) is absorbed at radius $r_{SW}$. Outgoing longwave radiation (OLR) is emitted from the radius $r_{LW}$. The mass flux $\Phi$ is driven by an energy flux $(g_sr_s + L)\Phi$. The mass flux obviously originates from the surface, but it does not reach large speeds ($\sim w_c$) until it is close to the critical radius $r_c$. For the parameter space regime we will consider, this region is far enough above the radiative surfaces that the outflow near the radiative surfaces is negligible (Appendix \ref{sec:app}). This allows radiative transfer and hydrodynamic escape to be treated separately .}
			\label{fig:schematic}
		\end{center}
		
	\end{figure*}

	A schematic of our model is shown in Figure \ref{fig:schematic}. The waterworld gains energy from incoming stellar radiation, and loses it via longwave emission and atmospheric escape. We describe this by the following equation (\ref{energybalance}), which is similar to those of \cite{pierrehumbert10} and \cite{lehmer17}. However, we specifically account for radiative surface expansion in both the shortwave and longwave regimes:
	\begin{equation}
	\left(\frac{r_{SW}}{r_s}\right)^2\frac{1}{4}(1-A)F_{\textit{stel}} - (g_s r_s + L)\Phi = \left(\frac{r_{LW}}{r_s}\right)^2F_{\textit{out}}
	\label{energybalance}
	\end{equation}
	Here $A$ is the planetary albedo, $F_{stel}$ and $F_{out}$ are the incoming shortwave and outgoing longwave fluxes, respectively, $g_s$ and $r_s$ are the surface gravity and radius, respectively, $L$ is the latent heat of vaporization of water, $\Phi$ is the escaping mass flux, and $r_{SW}$ and $r_{LW}$ are, respectively the radii of the shortwave absorption and longwave emission surfaces. The critical radius $r_c$, where the outflow reaches the speed of sound $w_c$, is far enough above the radiative surfaces that the outflow near the radiative surfaces is negligible. Therefore, radiative transfer and hydrodynamic escape can be treated separately (Appendix \ref{sec:app}). In the low-gravity limit, the energy balance equation is consistent with expressions for volatile loss from comets \citep{lebofsky75,weissman80}: since there is no atmosphere, the scaling factors disappear, the outgoing flux reduces to the Stefan-Boltzmann law ($F_{out} = \sigma T_s^4$), and the gravity term ($g_sr_s$) becomes negligible. Conversely, in the high gravity limit, the scaling factors and escape terms are negligible, resulting in a standard runaway greenhouse scenario.
	
	Here we obtain $r_{SW},r_{LW},F_{out}$ from line-by-line radiative transfer calculations (Section \ref{sec:rt}), and calculate $\Phi$ analytically, assuming hydrodynamic escape (Section \ref{sec:hde}). Then, \eqref{energybalance} is solved numerically in instellation-gravity space.
	
	\subsection{Radiative transfer}
	\label{sec:rt}
	To calculate outgoing longwave radiation and the radii of the radiative surfaces, we use the one-dimensional line-by-line radiative-convective model used previously by \cite{schaefer16} and \cite{wordsworth17}. As in \cite{schaefer16}, this model uses a truncated version of the HITEMP2010 line list \citep{rothman10} accurate below 1000 K. We assume that the atmosphere is in phase equilibrium with the surface reservoir and therefore the temperature profile is given by the saturation vapor pressure curve of water. We limit ourselves to clear-sky calculations, accounting for cloud albedo forcing using the standard assumption \citep{goldblatt13,goldblatt15} of an artificially large surface albedo (0.2). The potential effects of other cloud processes are discussed in Section \ref{sec:haboc}.
	
The ``radiative scaling factors" from \eqref{energybalance}, $(\frac{r_{SW}}{r_s})^2$ and $(\frac{r_{LW}}{r_s})^2$, are calculated using the method of \cite{goldblatt15}; we present a summary here. To a good approximation, emission of outgoing longwave radiation and absorption of incoming shortwave radiation occur at the altitude where optical depth $\tau=1$, measured from the top of the atmosphere. These altitudes can be calculated from the output of the radiative transfer model for a given surface gravity $g_s$ and temperature profile using the equation of hydrostatic balance (taking into account the decrease in gravity with altitude, which is non-negligible for small worlds):
	\begin{equation}
	\frac{dp}{dz}= -\rho g_s \left(\frac{r_s}{r_s + z}\right)^2
	\end{equation}
	Here $r_s$ is the surface radius, $z$ the altitude, $p$ pressure, and $\rho$ mass density. Even for the most rapidly escaping atmospheres we consider, the velocity of the outflow near the radiative surface is small, such that hydrostatic balance can be assumed (Appendix \ref{sec:app}). Once the heights of unity optical depth as a function of wavenumber (spectral heights) $z_{\tau=1}(\nu)$ are obtained, the height of the radiative surface is calculated by a weighted average of $z_{\tau=1}(\nu)$ using the relevant irradiance $I(\nu)$. 
	\begin{equation}
	z_{\text{radiative surface}} = \frac{\int I(\nu) \times z_{\tau =1}(\nu) d\nu}{\int I(\nu) d\nu}
	\end{equation}
	For the longwave emission surface, $I(\nu)$ is the outgoing longwave spectral irradiance given by the radiative transfer calculation. For the shortwave emission surface, it is the Planck blackbody spectral irradiance (hereafter, Planck curve) centered around the stellar effective emission temperature. The effect of deviations from blackbody behavior in real stellar spectra is neglected here.
	
\begin{figure*}[!ht]
		\begin{center}
			\includegraphics[width=0.95\linewidth]{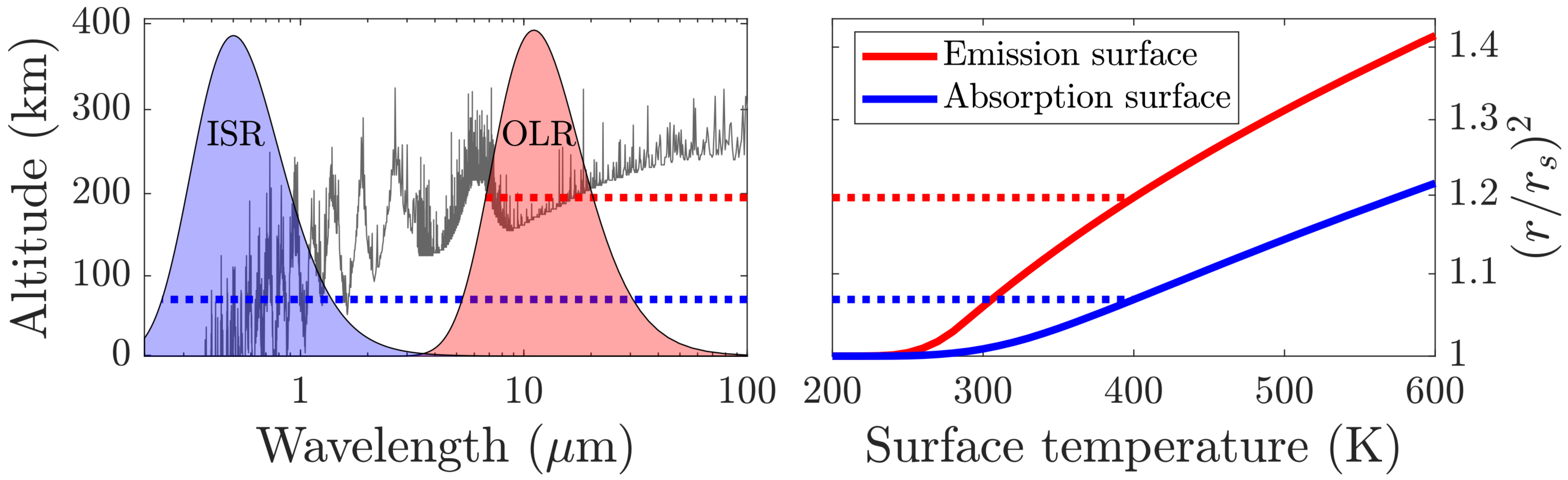}
			\caption{Left: spectral heights of $\tau = 1$ from the radiative transfer model, for a body with mass 0.12 $M_{\oplus}$ and a surface temperature of 400 K.  Right: ``radiative scaling factors" for the same body, as a function of temperature. The longwave scaling factor dominates the shortwave one, which is critical to the rest of this study. 
			The effective emission altitudes are determined by a weighted average of the spectral height where $\tau = 1$ over the irradiance output by the radiative transfer calculation and a Planck curve centered at the stellar emission temperature respectively. To illustrate this, scaled Planck curves for emission at 260 K (the temperature at the emission surface), and 5772 K (a Sun-like star) are overlaid.	The dashed lines show how averaging the altitudes over the different irradiances is responsible for the difference in the shortwave and longwave scaling factors. The scaling factors and spectral altitude of $\tau=1$ shown here are comparable to the Mars-like scenario in \cite{goldblatt15}. }
			\label{fig:tauscale}
		\end{center}
	\end{figure*}

	The nature of the water vapor absorption spectrum is such that $z_{\tau=1}(\nu)$ shows a decreasing trend with increasing $\nu$. Therefore, this averaging process means that the emission surface is always above the absorption surface. Therefore, the longwave ``scaling factor" is larger than its shortwave equivalent; indeed, their separation grows with temperature. This effect is critical to the rest of the study and is illustrated in Figure \ref{fig:tauscale}. The spectral heights of $\tau=1$ and the scaling factors produced by our model are in good agreement with those of \cite{goldblatt15}, who considered worlds with $0.12 M_{\oplus} < M < 10 M_{\oplus}$.
	\subsection{Hydrodynamic escape}
	\label{sec:hde}
	The problem of formulating the hydrodynamic escape from small bodies has been investigated previously. \citep{kuramoto94,pierrehumbert10,catling17,lehmer17,zahnle17}. Here we simplify the analysis by restricting ourselves to an isothermal escape formulation using the surface temperature $T_s$. This is because it is simple, easy to work with, and provides an upper bound on the escape rate. We are ultimately interested in water lifetime and habitability, and isothermal escape will provide a conservative estimate for both.
	
	The isothermal sound speed is $\sqrt{\frac{kT_s}{m}}\equiv w_c$, where $k$ is Boltzmann's constant and $m$ the molecular mass of water. Assuming transonic escape, there is a critical radius $r_c$ at which this speed is first reached. The outflow in spherically symmetric steady state is described by the continuity equation
	\begin{equation}
	\frac{d}{dr}(r^2 \rho w) = 0
	\end{equation}
	and the momentum conservation equation
	\begin{equation}
	w \frac{dw}{dr}  = - \frac{1}{\rho}\frac{dp}{dr} -\frac{GM}{r^2}
	\end{equation}
	
	where $\rho$ is density, $w$ is vertical velocity, $p$ is pressure, $G$ is Newton's gravitational constant, and $M$ is the mass of the body. These are combined (e.g. \citepl{catling17}, \citepl{lehmer17}) to form the isothermal planetary wind equation
	\begin{equation}
	(w^2 - w_c^2)\frac{1}{w}\frac{dw}{dr} = \frac{2w_c^2}{r}-\frac{GM}{r^2}
	\label{isoplawind}
	\end{equation}
	where the ideal gas law has also been used. At the critical radius $w=w_c$  and \eqref{isoplawind} yields 
	\begin{equation}
	r_c = \frac{GM}{2w_c^2}
	\end{equation} 
	Typically, $r_c \gtrsim 10 r_s$. Following \cite{lehmer17}, we integrate \eqref{isoplawind} from $r_s$ to $r_c$ and neglect the small kinetic energy term near the surface. This yields the mass flux, $\Phi$:
	\begin{equation}
	\Phi = \rho_s w_s = \rho_s w_c \left(\frac{r_c}{r_s}\right)^2 \text{exp}\left(-\frac{1}{2} + \frac{GM}{w_c^2}\left(\frac{1}{r_c}-\frac{1}{r_s}\right)\right)
	\label{massflux}
	\end{equation}
	$\Phi$ is a function of $T_s$ via $w_c$. The rate of mass loss is calculated as
	\begin{equation}
	\dot{M} = 4\pi r_s^2 \Phi
	\label{masslossrate}
	\end{equation}
	and the water lifetime $\tau_{water}$ is given by 
	\begin{equation}
	\tau_{water} = \frac{
		M_{water}}{\dot{M}}
	\label{waterlifetime}
	\end{equation}
	
	\section{Results}
	From the radiative transfer calculations, we generate outgoing radiation and scaling factors over a $g-T_s$ grid. For a given stellar flux $F_{stel}$ and gravity $g$, we can now numerically solve \eqref{energybalance} for $T_s$, the steady state surface temperature.  The steady-state surface temperature further determines escape rates and water lifetimes, which form the basis of the results presented here. 
	\subsection{No runaway greenhouse}
	\label{sec:norg}
	Terrestrial-mass worlds can experience the phenomenon known as the ``runaway greenhouse" \citep{komabayasi67,ingersoll69, kasting88, nakajima92, goldblatt12}: at high enough surface temperatures, total emission converges towards a limit. If the incoming stellar flux exceeds this threshold, there will be a runaway surface temperature increase, boiling off the oceans - making the world uninhabitable. Therefore, the runaway greenhouse is believed to set the ultimate inner edge of the habitable zone. \citep{kasting93,goldblatt12,kopparapu13,goldblatt13}
	
	Based on \eqref{energybalance} and the preceding analysis, it is clear that low-gravity waterworlds should behave differently. Although the flux per unit area from the effective emission surface will still reach a limit as temperature increases, the emission surface itself can keep gaining area relative to the absorption surface (Figure \ref{fig:tauscale}). If the ratio between the longwave and shortwave scaling factors does not reach a limit before escape becomes important, a runaway greenhouse limit will not be reached. 
	
\begin{figure}[h]
		\begin{center}
			\includegraphics[width=0.93\linewidth]{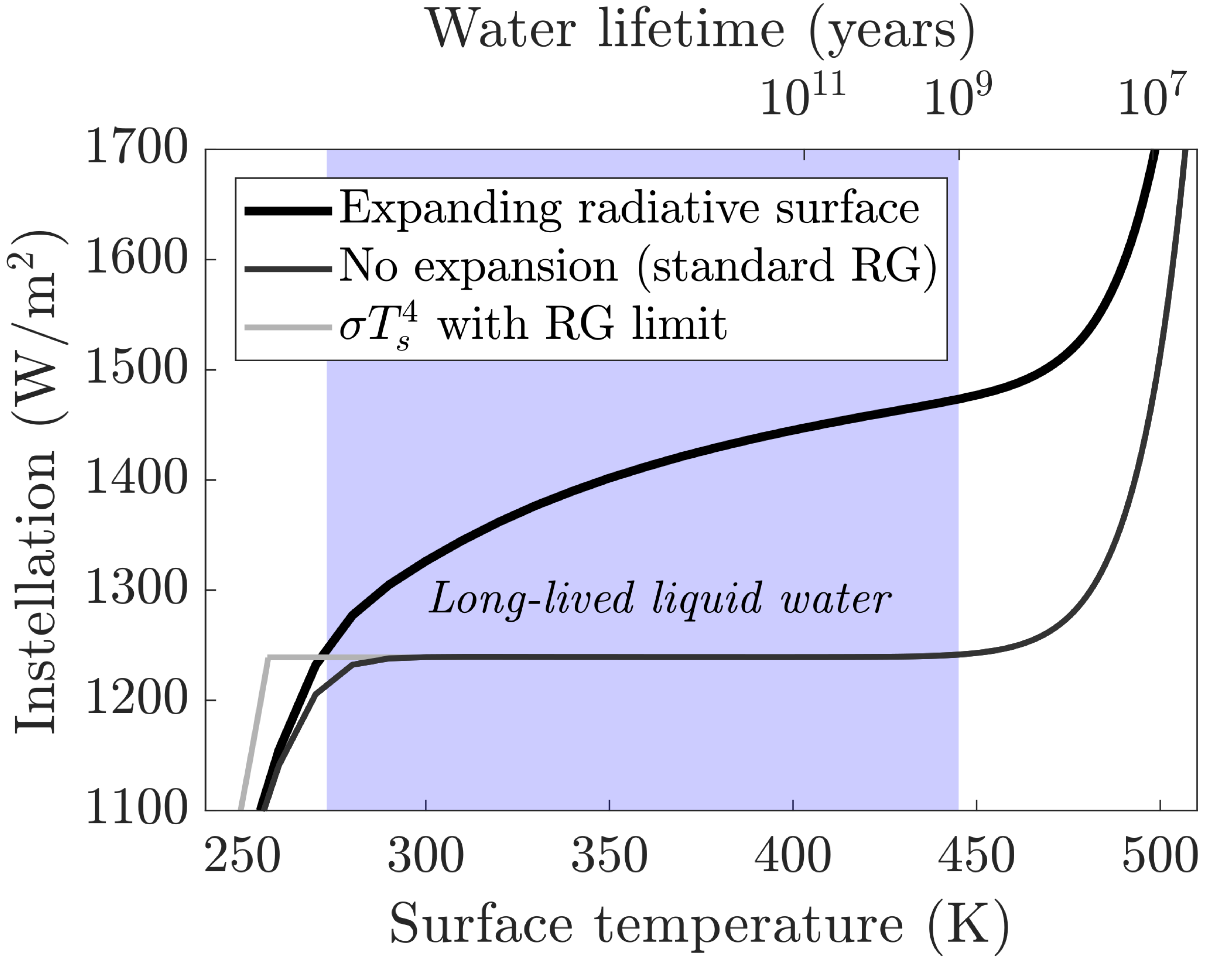}
			\caption{Steady state instellation \eqref{instab} as a function of surface temperature, for a world with $g = 2.3$ ms$^{-2}$ ($M\simeq0.08M_{\oplus}$). Long-lived surface liquid water exists for surface temperatures greater than 273 K and weak enough escape rates; this is denoted by the blue shaded region. We consider three differing radiation assumptions here: \textbf{1.} Line-by-line radiative transfer with radiative surface expansion taken into account (the model developed in this study). The steady-state instellation does not reach a limit before escape becomes important; thus, the runaway greenhouse is suppressed. \textbf{2.} Line-by-line radiative transfer without radiative surface expansion taken into account. This corresponds to the standard 'runaway greenhouse' scenario. \textbf{3.} $\sigma T_s^4$ flux from the surface until the runaway greenhouse limit. This corresponds with the assumptions made by \cite{lehmer17} for Europa- and Ganymede-like worlds. \textbf{1} predicts by far the widest habitable zone, showing the importance of radiative surface expansion. In contrast, \textbf{2} and \textbf{3} would predict a very narrow and a non-existent habitable zone, respectively.}
			\label{fig:norg}
		\end{center}
	\end{figure}

	We investigate this quantitatively by studying the steady-state instellation $F_{stel}$ as a function of surface temperature. \eqref{energybalance} can be rewritten in the form
	\begin{equation}
	F_{\textit{stel}}  = \frac{ (\frac{r_{LW}}{r_s})^2F_{\textit{out}} + (g_s r_s + L)\Phi}{(\frac{r_{SW}}{r_s})^2\frac{1}{4}(1-A)}
	\label{instab}
	\end{equation}
	$F_{stel}(T)$ is plotted in Figure \ref{fig:norg}, for a given $g=2.3$ ms$^{-2}$, corresponding to $M\simeq0.08 M_{\oplus}$. This value is illustrative because the escape rate increases slowly enough with $T$ that the radiative surface expansion effect can be clearly observed.  Water lifetimes are also added: they are a nonlinear function of temperature. For comparison, we also plot the results of two other radiation assumptions: the same line-by-line calculation with radiative surface expansion ignored, and blackbody emission from the surface ($\sigma T_s^4$) until the runaway greenhouse limit is reached. The latter approach was used by \cite{lehmer17}, to study Europa- and Ganymede-like worlds. Hydrodynamic escape is assumed to be governed by \eqref{massflux} in all cases. This is responsible for the rapid increase in $F_{stel}(T)$ as higher temperatures are approached. We note that such  states with high mass-flux are ``quasi-steady": the rapid expulsion of water into space is not sustainable indefinitely.
	
	A given steady-state instellation will allow for a habitable (long-lived liquid water) state if the surface temperature is larger than 273 K and the water inventory is long-lived enough. In Figure \ref{fig:norg} and the rest of this study we choose $10^9$ years as a cut-off; this choice is further discussed in Section \ref{sec:hz}. In our model, where radiative surface expansion is considered, $F_{stel}(T)$ never reaches a limit. Therefore, there is no runaway greenhouse. This leads to a relatively wide habitable zone. In contrast, neglecting radiative surface expansion but still using the line-by-line model would predict an extremely thin habitable zone, and assuming blackbody emission until the runaway greenhouse would predict no habitable zone at all.
	
	\subsection{Long-lived liquid water habitable zone}
	\label{sec:hz}
	We now expand the picture of Section \ref{sec:norg} and Figure \ref{fig:norg} into gravity space, allowing the instellation to be set by the orbital distance. We choose a water lifetime of $10^9$ years as an inner habitability boundary. While the principal requirement for the water lifetime habitability boundary is that it feasibly allows the evolution of life, this particular choice of timescale also ensures that such worlds have a reasonable chance of being habitable at the time of observation - even when the system's age is uncertain. With this definition, the ``longer-lived states" in our model consistently have surface temperatures below proposed upper bounds for survival of life \citep{bains15}, suggesting that temperature is not additionally prohibitive. 
	
\begin{figure*}[!ht]
		\begin{center}
			\includegraphics[width=\textwidth]{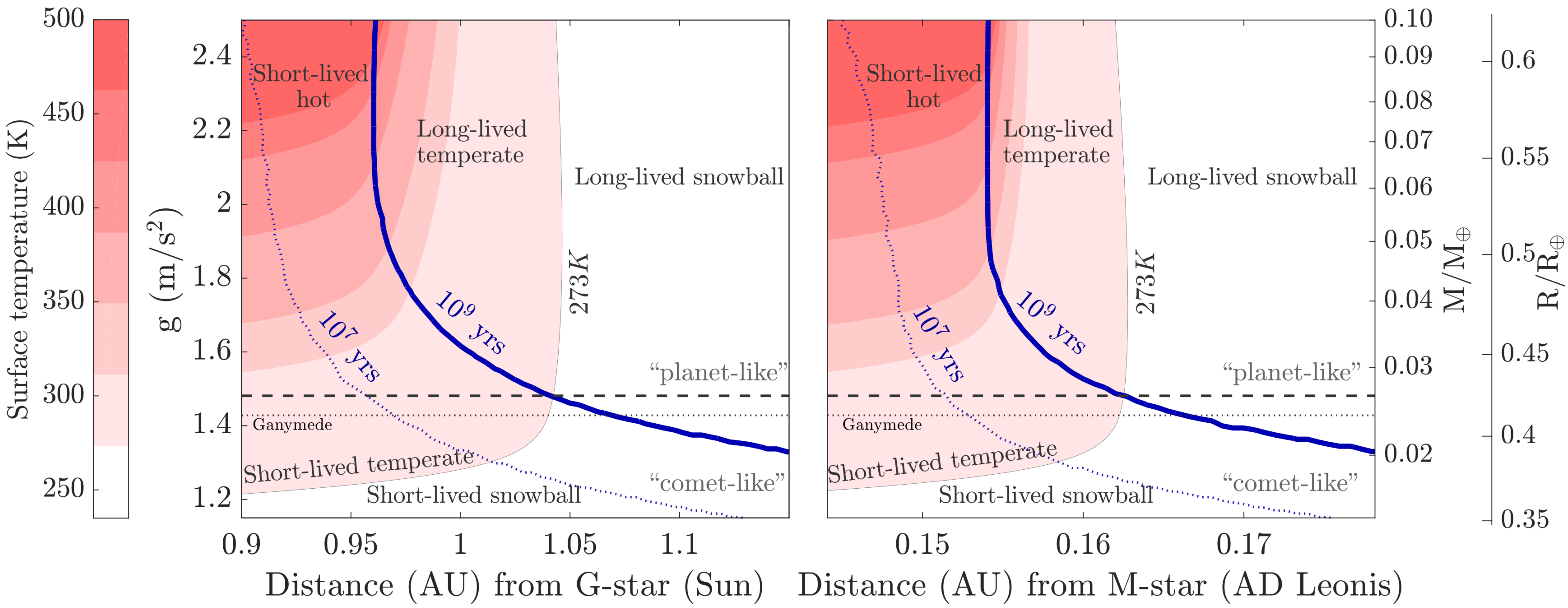}
			\caption{Steady-state surface temperatures and water lifetimes, obtained by numerically solving \eqref{energybalance}. We consider two cases: worlds orbiting a G-star (modeled on the Sun), and worlds orbiting an M-star (modeled on AD Leonis). A range of climate states is possible: short-lived hot, temperate, or snowball, and long-lived temperate or snowball. A conservative estimate for the ``habitable zone" (neglecting possible geochemical cycling) is defined simply as the region enclosed by the 273 K temperature contour and the $10^9$ year water lifetime contour. The width of this habitable zone decreases with gravity, ultimately disappearing completely - this is the transition from ``planet-like" to ``comet-like", marked by the black dashed line. For worlds whose mass-radius relationship is accurately described by \eqref{sotin}, this boundary is at $g\simeq 1.48$ ms$^{-2}$, or $M/M_{\oplus} \simeq 0.0268$. The mass of Ganymede is denoted by the dotted line; we see that it conservatively classifies as ``comet-like". However, as the $10^7$ year water lifetime contour (blue dashed line) shows, the habitable zone can extend both spatially and to lower masses if the lifetime criterion is relaxed. The estimated transition from planet-like to comet-like would likewise move to lower masses. }
			\label{fig:wl}
		\end{center}
	\end{figure*}

	As can be observed from Figure \ref{fig:tauscale}, stellar emission temperature plays a role in determining the size of the shortwave scaling factor. We thus consider two limiting cases: an M-star (modeled on AD Leonis) and a G-star (modeled on the Sun). The resulting steady-state temperatures and the $10^9$ year water lifetime contour are shown in Figure \ref{fig:wl}. We label five qualitatively different climate states: long-lived snowball, long-lived temperate, short-lived snowball, short-lived temperate and short-lived hot. The mass of Ganymede and a $10^7$ year water lifetime contour are plotted additionally for reference.
	
	The ``habitable zone" is defined by the area enclosed by the 273 K temperature contour and the $10^9$ year water lifetime contour. The conservative maximum width in the low-gravity regime is $\sim 0.08$ AU for G-stars and $\sim 0.008$ AU for M-stars. Relaxing the $10^9$ year habitability boundary would extend the habitable zone both spatially and to lower gravities. Worlds with appropriate geochemical cycling could of course have a wider habitable zone, because the outer edge would move outwards.

	\subsection{The planet-comet transition}
	\label{sec:planetcomet}
	The width of the habitable zone shown in Figure \ref{fig:wl} decreases with gravity, ultimately disappearing. This marks the transition from ``planet-like" (habitable at some stellar flux) to ``comet-like" (never habitable regardless of stellar flux). For worlds with a mass-radius scaling accurately described by \eqref{sotin}, our model conservatively gives this boundary at $g\simeq 1.48$ ms$^{-2}$, or $M/M_{\oplus} \simeq 0.0268$.  The boundary is essentially independent of stellar spectral type, because the differences in the shortwave scaling factor are negligible for waterworld surface temperatures near 273 K. 
	
	\subsection{Stellar flux-driven deglaciation}
	\label{sec:hysteresis}
\begin{figure}[h]
		\begin{center}
			\includegraphics[width=\linewidth]{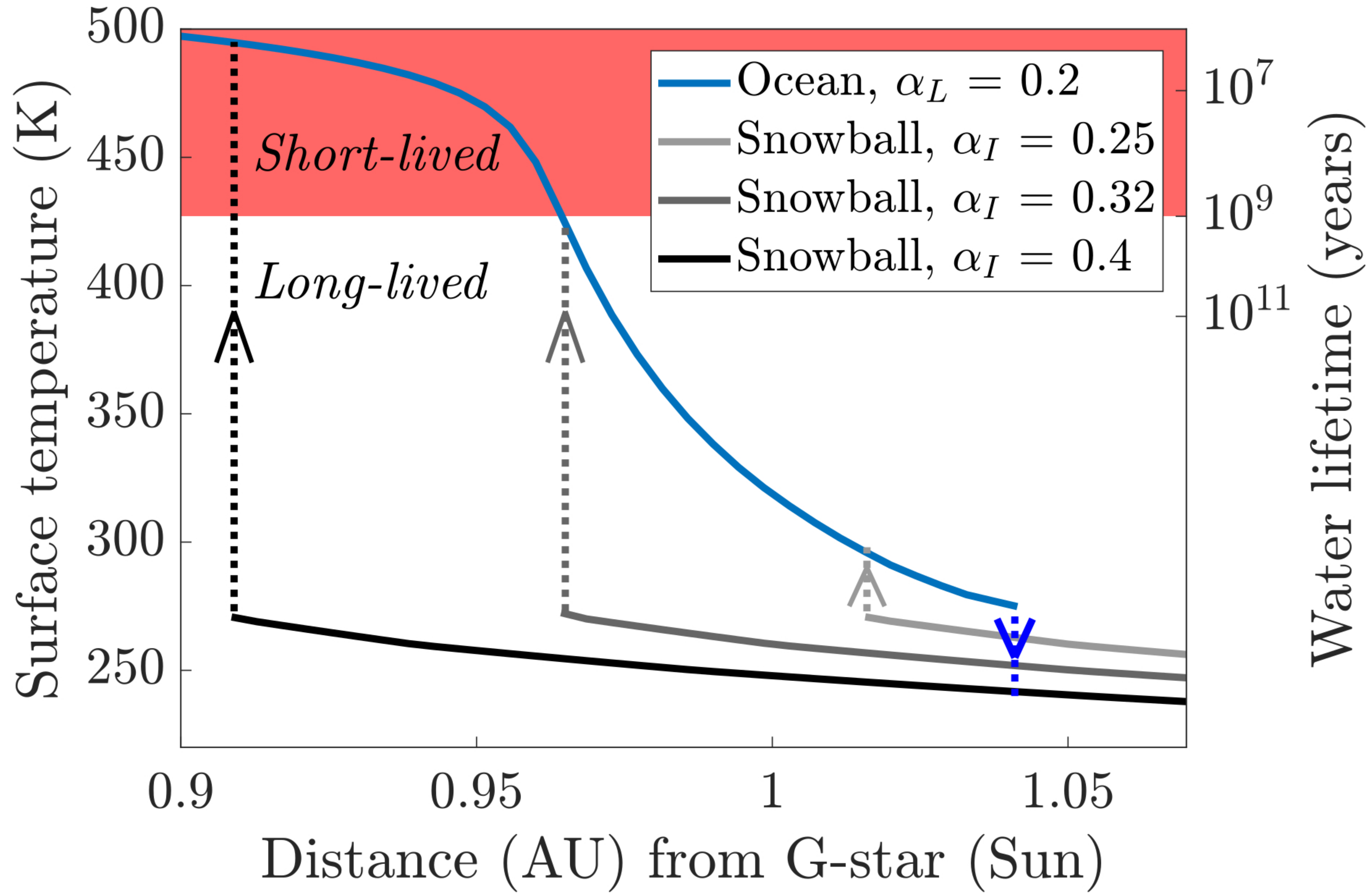}
			\caption{Hysteresis plot, for $g=2.3$ m~s$^{-2}$ ($M\simeq0.08M_{\oplus}$) and a sun-like star. We observe that a snowball state (``cold history") experiencing a stellar flux-driven deglaciation generally bypasses the long-lived habitable state entirely, except for when $\alpha_I$ values are very low.}
			\label{fig:hyst}
		\end{center}
	\end{figure}

	Although our model predicts a habitable climate state, such a state is not necessarily easy to access. Figure \ref{fig:wl} exhibits a range of possible climate states: short-lived hot, temperate or snowball, and long-lived temperate or snowball. The ice-albedo feedback can hamper transitions from snowball states to temperate states: this has already been demonstrated for terrestrial-mass worlds \citep{yang17}. We can incorporate the ice-albedo feedback into our model using the simple albedo step function
	\begin{equation}
	A(T_s) = \begin{cases} 
	\alpha_I & T_s < 273 \text{ K}\\
	\alpha_L & T_s \geq 273 \text{ K} 
	\end{cases}
	\label{albedo}
	\end{equation}
	where $\alpha_I$ is the albedo of the icy (snowball) state, and $\alpha_L$ is the albedo when there is surface liquid water. Hysteresis plots for different choices of $\alpha_I$ are shown in Figure \ref{fig:hyst}. We observe that a snowball state experiencing a stellar flux-driven deglaciation generally bypasses the long-lived state entirely, except for very low $\alpha_I$ values. Although the mechanism setting the inner edge of the habitable zone is different, the conclusion of the habitable state likely being bypassed upon stellar flux-driven deglaciation is the same as that of \cite{yang17}.
	
	\section{Discussion}
	\label{sec:disc}
	\subsection{Habitability: other considerations}
	\label{sec:haboc}
	Beyond those considered in this study, other factors play a role in determining the habitability of low-gravity waterworlds. For example, we have restricted ourselves to pure water vapor atmospheres; in reality, a variety of species could be present, depending on interior and atmospheric chemistry. 	We did not consider the potential effects of clouds except via an increased planetary albedo. Cloud greenhouse forcing could reduce outgoing radiation as it does for terrestrial-mass worlds, and clouds in an expanded low-gravity atmosphere could increase absorbed flux by scattering more stellar radiation towards the surface. Depending on whether cloud effects contribute more to cooling or warming, the size of the habitable zone presented in this study could either increase or decrease. The net effect should be carefully assessed in the future using three-dimensional models (e.g. \citeauthor{pierrehumbert16} \citeyear{pierrehumbert16}). Finally, as in our solar system, low-gravity waterworlds may often be found as exomoons around giant planets. Then, tidal forces are important for habitability \citep{heller13}, becoming especially deleterious around M-stars \citep{zollinger17}.
	
	Even if a habitable state exists, it is not necessarily ``accessible", as we showed in Section \ref{sec:hysteresis}. Worlds with a warm history have access to the entire habitable zone derived here; the proto-atmosphere phenomenon discussed by \cite{kuramoto94} could make a warm start more likely. However, for worlds with a cold history, stellar-flux-driven deglaciation into a long-lived state only occurs for artificially low ice albedo values. Somewhat more favorable conditions might be found around M-stars, where the ice-albedo feedback is weakened \citep{joshi12, shields13}. However, such systems experience another problem: the dramatic pre-main sequence evolution of the star may strip potentially habitable worlds of their water early on \citep{ramirez14,luger15,tian15}. Further research is required to integrate all of these considerations.
	
	In this study we have used the conventional but limited definition of habitability as the ability to maintain surface liquid water on sufficient timescales. While our model categorizes worlds of Europa's or Ganymede's size as ``comet-like" (unable to maintain surface liquid water for Gyr timescales regardless of stellar flux), that does not necessarily make them uninhabitable. For example, there is a large body of work on how life could survive on Europa within a subsurface ocean, supported by energy from a geologically active interior and favorable chemistry \citep{reynolds83,mccollom99,gaidos99,chyba00}. Detection of such life remotely would be extremely challenging. Thus the question of life on ``comet-like" worlds ultimately remains open, if unanswerable at present.
	
	\subsection{Observing low-gravity waterworlds}
	Observation of low-gravity waterworlds outside of our solar system is possible in principle. The smallest exoplanet discovered to date is the rocky Kepler 37-b, with a radius of $0.303R_{\oplus}$ \citep{barclay13}. This is already smaller than all of the low-gravity waterworlds modeled in this study. Regarding exomoons, a previous search using the Kepler Space Telescope \citep{kipping15} was sensitive to masses as low as $1.7$ $M_{\text{Ganymede}}\simeq 0.043$ $M_{\oplus}$, which includes most of the mass range for habitable low-gravity waterworlds shown in Figure \ref{fig:wl}. Unfortunately, due to the nature of transit photometry, low-gravity worlds can currently only be found in orbits with relatively small semi-major axes; this is likely to preclude long-lived water. Correctly classifying waterworlds presents another issue: under current observational uncertainties in this size regime \citep{jontofhutter15}, mass-density determinations alone are likely to be insufficient \citep{sotin07}.
	
	The phenomena described in this study will have observable effects. Radiative surface expansion increases transit depths; planetary density will be underestimated by a factor of $r_{SW}^3/r^3$, a decrease of up to $15\%$ for temperate worlds. Clouds or escaping dust will amplify this: the latter already provides a means to observe the much smaller exo-comets \citep{kiefer14, rappaport17}, and has also been proposed to explain anomalously low-density sub-Neptunes \citep{wang19}. Escaping H atoms could produce a detectable Ly-$\alpha$ signal; for exomoons, these atoms would likely accumulate in a torus around the host planet \citep{lehmer17}. Transmission spectroscopy has already been successfully used to study Earth-mass planets (e.g. \citeauthor{dewit16} \citeyear{dewit16}); for low-gravity waterworlds, the highly expanded atmosphere will amplify the variations in such a spectrum. Coupling the model developed here with a forward model for observables would be an interesting follow-up to this study, and would aid future detection efforts.\\
	
	\subsection{Future directions}
	\label{sec:fd}
	Further work could consider more complicated models of hydrodynamic escape. Since the isothermal escape approximation represents an upper bound, such work is likely to widen the habitable zone presented in Figure \ref{fig:wl}. 
	
	The two key phenomena governing evolution on low-gravity waterworlds, radiative surface expansion and hydrodynamic escape, are treated separately in our model. This is possible while the outflow near the radiative surface is negligible, which is the case for the parameter space regime we consider. While our model is sufficient to estimate constraints on habitability, the scenario of non-negligible outflow near the radiative surface remains of physical interest, and should be explored further.
	
	Finally, the planet-comet transition calculated in this study represents a hard lower gravity boundary for habitability in the conventional sense. An upper gravity boundary is likely given by the onset of runaway gas accretion, which distinguishes super-Earths from gas giants (e.g. \citeauthor{ginzburg16} \citeyear{ginzburg16}). Such boundaries, when used in tandem, could be a useful tool when estimating a total number of potentially habitable worlds.

	\section{Conclusion}
	\label{sec:conc}
	Low-gravity waterworlds experience radiative surface expansion \citep{goldblatt15} and hydrodynamic escape \citep{kuramoto94, pierrehumbert10, catling17, zahnle17, lehmer17}. This study is the first to explore their combined effect on long-term atmospheric evolution and habitability. We showed that radiative surface expansion suppresses the runaway greenhouse (Section \ref{sec:norg}), the existence and location of a surface liquid water habitable zone whose inner edge is set by escape (Section \ref{sec:hz}), and that the ice-albedo feedback works against stellar-flux-driven deglaciations to long-lived states (Section \ref{sec:hysteresis}).  Combining these effects has allowed us for the first time to calculate the transition from ``planet-like" to ``comet-like" regimes for the evolution of a world's H$_2$O inventory (Section \ref{sec:planetcomet}). Comet-like worlds cannot sustain surface liquid water on Gyr timescales. Therefore, by the criterion of long-lived surface liquid water, this provides a hard lower boundary on habitability in terms of gravity. 
	
	\acknowledgements C. W. A. acknowledges support from the Harvard Origins of Life Initiative, and thanks E. Kite and P. Niraula for helpful discussions. This work was supported by NASA Habitable Worlds grant NNX16AR86G.
	
	\appendix
	\section{Hydrostatic balance at the radiative surface}
	\label{sec:app}
	As stated in Section \ref{sec:eb} and Section \ref{sec:rt}, our model can treat radiative surface expansion and hydrodynamic escape separately because the velocity of the outflow at and below the radiative surface is negligible.  Specifically, it is small enough to allow a calculation of the radiative surface height using the hydrostatic relation (Section \ref{sec:rt}). Here, we make this argument quantitatively.
	
\begin{figure}[h]
		\begin{center}
			\includegraphics[width=0.9\linewidth]{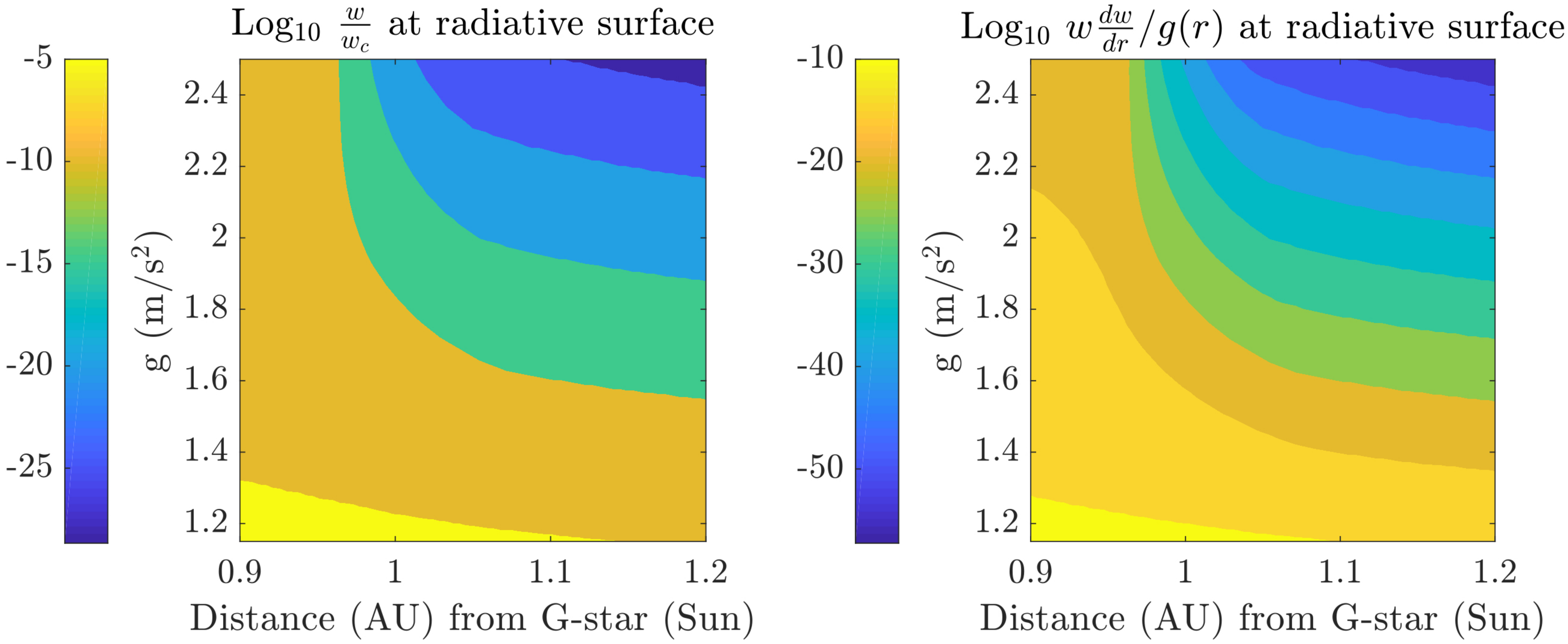}
			\caption{Evaluating the sizes of $w/w_c$ and $w \di{w}{r}/g$ at the longwave radiative surface, using equations \eqref{wanalytic} and \eqref{wdwdranalytic}. Since $w/w_c \ll 1$  at the longwave radiative surface, the use of \eqref{wdwdranalytic} to calculate $w \di{w}{r}/g$ is justified. Since $w \di{w}{r}/g \ll 1$ at the longwave radiative surface, the use of the hydrostatic approximation in Section \ref{sec:rt} is justified. }
			\label{fig:wdwdr}
		\end{center}
	\end{figure}

	For spherically symmetric hydrodynamic escape, the most complete momentum equation is
	\begin{equation}
	\rho w\di{w}{r} + \di{p}{z} = -\rho g
	\end{equation} 
	where the gravity $g$, pressure $p$, outflow $w$ are all functions of $r$. This is accurately approximated by a hydrostatic relation $\di{p}{z} = - \rho g$ if the inertial term $\rho w \di{w}{r}$ is sufficiently small. Thus it suffices to check whether the dimensionless number
	\begin{equation}
	|w \di{w}{r}|/|g(r)|
	\end{equation}
	is much smaller than 1 at and below the (longwave) radiative surface. We can obtain an analytical expression for $w \di{w}{r}$ by considering an integrated solar wind equation for isothermal escape \citep{parker63,catling17}:
	\begin{equation}
	\ln(\frac{w}{w_c})-\frac{1}{2}(\frac{w}{w_c})^2 = -\frac{2r_c}{r} - 2 \ln (\frac{r}{r_c}) + \frac{3}{2}
	\end{equation}
	For $w/w_c \lesssim 0.4$, it is a good approximation to drop the $w^2$ term and write
	\begin{equation}
	w(r) = w_c (\frac{r_c}{r})^2 \text{exp}(\frac{3}{2} - 2\frac{r_c}{r})
	\label{wanalytic}
	\end{equation}
	Hence, 
	\begin{equation}
	w\di{w}{r} = 2w_c^2(\frac{r_c^5}{r^6}- \frac{r_c^4}{r^5})\text{exp}(3 - 4\frac{r_c}{r})
	\label{wdwdranalytic}
	\end{equation}
	Across the parameter range presented in Figure \ref{fig:wl}, using the steady state temperatures, we can thus calculate $w$ and $w\di{w}{r}$ at the radiative surface using \eqref{wanalytic} and \eqref{wdwdranalytic}. In this regime, it is the case that $w\di{w}{r}$ monotonically increases until the radiative surface. Therefore, when considering the accuracy of the hydrostatic approximation in calculating the radiative surface height, it is ultimately sufficient to evaluate $w \di{w}{r}/g$ at the longwave radiative surface, and verify that it is much smaller than 1. In Figure \ref{fig:wdwdr}, we plot logarithms of $w/w_c$ and $w \di{w}{r}/g$ at the longwave radiative surface, over the same $g$-distance regime covered by Figure \ref{fig:wl}, finding that both of these terms are much smaller than 1. Since $w/w_c \ll 1$  at the longwave radiative surface, the use of \eqref{wdwdranalytic} to calculate $w \di{w}{r}/g$ is justified. Since $w \di{w}{r}/g \ll 1$ at the longwave radiative surface, the use of the hydrostatic approximation in Section \ref{sec:rt} is justified. 


\begin{thebibliography}{}
		
		\bibitem[Abbot et al.(2012)]{abbot12} Abbot, D. S., Cowan, N. B., \& Ciesla, F. J. 2012. \apj, 756(2), 178.
		\bibitem[Bains et al.(2015)]{bains15} Bains, W., Xiao, Y., \& Yu, C.
		2015, Life, 5(2), 1054-1100.
		\bibitem[Barclay et al.(2013)]{barclay13} Barclay, T., Rowe, J. F., Lissauer, J. J., et al. 2013, Nature, 494(7438), 452.
		\bibitem[Catling and Kasting(2017)]{catling17} Catling, D. C., \& Kasting, J. F. 2017, Cambridge University Press
		\bibitem[Chela-Flores(2010)]{chelaflores10} Chela-Flores, J., 2010.  IJAsb, 9(2), 101-108.
		\bibitem[Chyba(2000)]{chyba00} Chyba, C. F. 2000. Nature, 403(6768), 381.
		\bibitem[Chyba and Phillips(2001)]{chyba01} Chyba, C. F., \& Phillips, C. B.,2001. PNAS, 98(3), 801-804.
		\bibitem[de Wit et al.(2016)]{dewit16} de Wit, J., Wakeford, H. R., Gillon, M., et al. 2016. Nature, 537(7618), 69.	
		\bibitem[Ding and Pierrehumbert(2018)]{ding18} Ding, F. \& Pierrehumbert, R. T., 2018. \apj, 867(1), p.54.
		\bibitem[Gaidos et al.(1999)]{gaidos99} Gaidos, E. J., Nealson, K. H., \& Kirschvink, J. L. 1999. Science, 284(5420), 1631-1633.
		\bibitem[Ginzburg et al.(2016)]{ginzburg16} Ginzburg, S., Schlichting, H. E., \& Sari, R. E. 2016. \apj, 825(1), 29.
		\bibitem[Goldblatt(2015)]{goldblatt15} Goldblatt, C.\ 2015, Astrobiology, 15, 5
		\bibitem[Goldblatt et al.(2013)]{goldblatt13} Goldblatt, C., Robinson, T. D., Zahnle, K. J., et al. 2013. Nature Geoscience, 6(8), 661.
		\bibitem[Goldblatt and Watson(2012)]{goldblatt12} Goldblatt, C., \& Watson, A. J. 2012, Phil. Trans. R. Soc. A, 370, 4197–4216
		\bibitem[Grasset et al.(2013)]{grasset13} Grasset, O., Dougherty, M. K., Coustenis, A., et al. 2013, \planss, 78, 1-21.
		\bibitem[Heller and Barnes(2013)]{heller13} Heller, R., \& Barnes, R. 2013, Astrobiology. 13, 1, 18-46
		\bibitem[Huang(1959)]{huang59} Huang, S. S. 1959. American Scientist, 47(3), 397-402.
		\bibitem[Ingersoll(1969)]{ingersoll69} Ingersoll, A. P. 1969. Journal of the Atmospheric Sciences, 26(6), 1191-1198.
		\bibitem[Jontof-Hutter et al.(2015)]{jontofhutter15} Jontof-Hutter, D., Rowe, J. F., Lissauer, J. J., Fabrycky, D. C., \& Ford, E. B.
		2015, Natur, 522, 321
		\bibitem[Joshi and Haberle(2012)]{joshi12} Joshi, M. M., \& Haberle, R. M. 2012, Astrobiology, 12, 3-8
		\bibitem[Kasting(1988)]{kasting88} Kasting, J. F. 1988. Icarus, 74(3), 472-494.
		\bibitem[Kasting et al.(1993)]{kasting93} Kasting, J.F., Whitmire, D.P. \& Reynolds, R.T., 1993. \icarus, 101(1), 108-128.
		\bibitem[Kiefer et al(2014)]{kiefer14} Kiefer, F., des Etangs, A. L., Augereau, J. C., et al. 2014. \aap, 561, L10.
		\bibitem[Kipping et al.(2015)]{kipping15} Kipping, D. M., Schmitt, A. R., Huang, X., et al. 2015. \apj, 813(1), 14.
		\bibitem[Kite and Ford(2018)]{kite18} Kite, E. S., \& Ford, E. B. 2018, \apj, 864, 1
		\bibitem[Komabayasi(1967)]{komabayasi67} Komabayasi, M. 1967. Journal of the Meteorological Society of Japan. Ser. II, 45(1), 137-139.
		\bibitem[Kopparapu et al.(2013)]{kopparapu13} Kopparapu, R. K., Ramirez, R., Kasting, J. F., et al. 2013. \apj, 765(2), p.131.
		\bibitem[Kopparapu et al.(2014)] {kopparapu14} Kopparapu, R. K., Ramirez, R. M., SchottelKotte, J., et al. 2014. \apjl, 787(2), L29.
		\bibitem[Kuramoto and Matsui(1994)]{kuramoto94} Kuramoto, K., \& Matsui, T. 1994, \jgr 21,183-21,200
		\bibitem[Lebofsky(1975)]{lebofsky75} Lebofsky, L. A. 1975. \icarus, 25(2), 205-217.
		\bibitem[Lehmer et al.(2017)]{lehmer17} Lehmer, O. R., Catling, D. C., \& Zahnle, K. J.\ 2017, \apj, 839, 1, 32
		\bibitem[Luger and Barnes(2015)]{luger15}  Luger, R., \& Barnes, R. 2015, Astrobiology, 15, 119-143
		\bibitem[McCollom(1999)]{mccollom99} McCollom, T. M. 1999. \jgr, 104(E12), 30729-30742.
		\bibitem[Nakajima et al.(1992)]{nakajima92} Nakajima, S., Hayashi, Y. Y., \& Abe, Y. 1992. Journal of the Atmospheric Sciences, 49(23), 2256-2266.
		\bibitem[Parker(1963)]{parker63} Parker, E. N., 1963. Interscience Publishers.
		\bibitem[Pierrehumbert(2010)]{pierrehumbert10} Pierrehumbert, R. T.\ 2010, Cambridge University Press
		\bibitem[Pierrehumbert and Ding(2016)]{pierrehumbert16} Pierrehumbert, R. T., \& Ding, F. 2016. Proc. Royal Soc. Lond. A, 472(2190), 20160107.
		\bibitem[Rappaport et al.(2017)]{rappaport17} Rappaport, S., Vanderburg, A., Jacobs, T., et al. 2017. \mnras, 474(2), 1453-1468.
		\bibitem[Ramirez(2018)]{ramirez18} Ramirez, R., 2018. Geosciences, 8(8), p.280.
		\bibitem[Ramirez and Kaltenegger(2014)]{ramirez14} Ramirez, R. M. \& Kaltenegger, L., 2014. \apjl, 797(2), p.L25.
		\bibitem[Ramirez and Levi(2018)]{ramirezlevi18} Ramirez, R. M. \& Levi, A., 2018. \mnras, 477(4), pp.4627-4640.
		\bibitem[Reynolds et al.(1983)]{reynolds83} Reynolds, R. T., Squyres, S. W., Colburn, D. S., et al. 1983. Icarus, 56(2), 246-254.
		\bibitem[Rothman et al.(2010)]{rothman10} Rothman, L. S., Gordon, I. E., Barber, R. J., et al. 2010. \jqsrt, 111(15), 2139-2150.
		\bibitem[Schaefer et al.(2016)]{schaefer16} Schaefer, L., Wordsworth, R. D., Berta-Thompson, Z., et al. 2016. \apj, 829(2), 63.
		\bibitem[Seager(2013)]{seager13} Seager, S., 2013. Science, 340(6132), pp.577-581.
		\bibitem[Shields et al.(2013)]{shields13} Shields, A. L., Meadows, V. S., Bitz, C. M., et al. 2013. Astrobiology, 13(8), 715-739.
		\bibitem[Sotin et al.(2007)]{sotin07} {Sotin}, C., Grasset, O.  \& Mocquet, A.\ 2007, \icarus, 191, 337-351
		\bibitem[Tian and Ida(2015)]{tian15} Tian, F., \& Ida, S. 2015, Nature Geoscience, 8, 177-179
		\bibitem[Wang and Dai(2019)]{wang19} Wang, L., \& Dai, F. 2019. \apjl,  873(1), p.L1.
		\bibitem[Weissman(1980)]{weissman80} Weissman, P. R. 1980. \aap, 85, 191-196.
		\bibitem[Wordsworth et al.(2017)]{wordsworth17} Wordsworth, R., Kalugina, Y., Lokshtanov, S., et al. 2017 \grl, 44
		\bibitem[Wordsworth and Pierrehumbert(2013)]{wordsworth13} Wordsworth, R. D. and Pierrehumbert, R. T., 2013. \apj, 778(2), p.154.
		\bibitem[Yang et al.(2017)]{yang17} Yang, J., Ding, F., Ramirez, R. M., et al. 2017. Nature Geoscience, 10, 556-560
		\bibitem[Zahnle and Catling(2017)]{zahnle17} Zahnle, K. J., \& Catling, D. C., 2017. \apj, 843, 122
		\bibitem[Zollinger et al.(2017)]{zollinger17} Zollinger, R. R., Armstrong, J. C., \& Heller, R. 2017. \mnras, 472(1), 8-25.
		
	\end{thebibliography}
\end{document}